\documentclass[a4paper,10pt]{article}
\usepackage[latin1]{inputenc}
\usepackage[fleqn]{amsmath}
\usepackage{graphicx}
\usepackage{epsfig}
\usepackage{float}
\usepackage{caption}
\usepackage{amsfonts}
\begin{document}

\begin{center}
{\Large\bf Dynamical systems study of Chameleon scalar field}
\\[15mm]
Nandan Roy \footnote{E-mail: nandan@iiserkol.ac.in} and
Narayan Banerjee \footnote{E-mail: narayan@iiserkol.ac.in}

{\em Department of Physical Sciences,~~\\Indian Institute of Science Education and Research-Kolkata,~~\\Mohanpur Campus-741252,India.}\\[15mm]
\end{center}

\begin{abstract}
The present work is an extensive study of the viable stable solutions of chameleon scalar field models leading to possibilities of an accelerated expansion of the universe. It is found that for various combinations of the chameleon field  potential $V(\phi)$ and the coupling $f(\phi)$ of the chameleon field with matter, a stable solution for an accelerated expansion is quite possible. The investigation provides a diagnostics for the stability criteria for all sorts of combinations of $V(\phi)$ and $f(\phi)$.
\end{abstract}


\section{Introduction:}
Although the observational evidences in favour of the recent accelerated expansion of the universe are mounting everyday, the driver of this acceleration has neither been detected nor does any of the candidates has a clear verdict on theoretical grounds. A cosmological constant is the most favoured one, but it has its share of problems \cite{varun, paddy, sami}. Amongst a host of other possibilities, a scalar field with a potential, leading to an effective negative pressure, popularly called a quintessence field, appears to enjoy the second highest priority \cite{martin}. Relativistic theories of gravitation, other than the General Theory of Relativity, are also invoked. For example, an $f(R)$ theory of gravity, where the Ricci scalar $R$ in the Einstein action is replaced by an analytical function $f=f(R)$, does well in explaining the accelerated expansion \cite{review}. Another popular direction is to look for an accelerated model in an already existing nonminimally coupled
  scalar field theory, particularly Brans-Dicke theory or some of the modifications of the theory \cite{nbdp}. In such theories, the scalar field or some function of the same is nonminimally coupled to the gravity sector in the action in the form $f(\phi)R$. Recently another form of a nonminimally coupled scalar field theory, where the scalar field is coupled to the matter sector rather than the geometry, has been proposed. A chameleon field, characterised by a ``chameleon mechanism'' is an example of such a field  nonminimally coupled to matter  \cite{justin1, justin2}. A consequence of this mechanism is that the mass of the scalar field is not a constant but rather changes with the ambient matter distribution. This change of properties of the field acts as a buffer in connection with the observational bounds set on the mass of the scalar fields coupled to the matter sector  \cite{mota1}. \\

Chameleon scalar fields have many interesting features. For instance, a strongly coupled chameleon has the possibility of being detected by carefully designed experiments \cite{mota2, shaw1}. Possible effect of the chameleon field on the cosmic microwave background with possible observational imprints has been estimated by Davis, Schelpe and Shaw \cite{shaw2} and that on the rotation curve of galaxies by Burikham and Panpanich \cite{buri}. One very attractive feature of a chameleon field is that if it is coupled to an electromagnetic field \cite{brax1} in addition to the fluid, the fine tuning of the initial conditions on the chameleon may be resolved to a large extent \cite{mota3}. The remarkable features of the chameleon field theories are comprehensively summarized by Khoury \cite{khoury}. \\

Brax {\it et al} used this chameleon field as a dark energy\cite{brax}. This kind of interaction between the dark matter and the dark energy was investigated in detail by Das, Coarsaniti and Khoury\cite{das} in the context of the present acceleration of the universe. It was also shown that with a chameleon field of this sort, it is quite possible to obtain a smooth transition from a decelerated to an accelerated expansion for the universe\cite{nbsdkg}.

The success of the chameleon mechanism in explaining the current accelerated expansion and its lucrative properties which open up the possibility to evade a fine tuning of initial conditions, and its possible observational imprints inevitably attracted a lot of attention. The possibility of a scalar field nonminimally coupled to gravity, such as the Brans-Dicke scalar field, acting as a chameleon was discussed by Das and Banerjee \cite{sdnb}. Brans-Dicke scalar field acting as a chameleon with an infrared cut-off as that in the holographic models was discussed by Setare and Jamil \cite{jamil}. The field profile of a chameleon was discussed by Tsujikawa, Tamaki and Tavakol \cite{tavakol}. \\

The aim of the present work is to thoroughly investigate the stability criteria of the chameleon models in a spatially homogeneous and isotropic cosmology. There are two arbitrary functions of the chameleon field $\phi$ to start with, namely $V=V(\phi)$ and $f=f(\phi)$. Here $V$ is the dark energy potential and $f$ determines the coupling of the chameleon field with the matter sector. We broadly classify the functions into two categories, exponential and non-exponential. So there are four combinations in all. We investigate the conditions for having a stable solution for the evolution for each of these categories. We find that there are possibilities of finding a stable evolution scenario where the universe may settle into a phase of accelerated expansion. However, if both $V$ and $f$ are exponential functions of $\phi$, the stability is very strongly dictated by the model parameters. In fact it is noted that in this latter case there is a possibility of a transient accelerat
 ion at the present epoch but the final stable configuration of the universe is that of a decelerated expansion.\\

The method taken up is the dynamical systems study. The field equations are written as an autonomous system and the fixed points are found out. A stable fixed point indicates a sink and thus marks the possible stable final configuration of the universe whereas an unstable fixed point, indicating a source, may describe the possible beginning of the evolution. Application of dynamical systems in cosmological problems, mostly for scalar field distributions, is already there in the literature \cite{gunzig}. For detailed discussions on some early work on such investigations, we refer to the monograph by Coley \cite{coley} (see also \cite{ellis}). 

The paper is organized as follows. Section 2 deals with a chameleon scalar field model in a spatially flat homogeneous and isotropic universe. In section 3 the system of equations given in section 2 are written as an autonomous system. This section also includes a brief discussion of the method of the stability analysis that is used in the present work. The actual stability analysis in the four categories as mentioned is given in section 4. The fifth section presents an example of a chameleon field where the chameleon mechanism together with the constraints imposed by laboratory experiments does not have any contradiction with the cosmological requirement of the decelerated expansion entering into an accelerated phase at a redshift of 0.74\cite{farook}. The sxith and the final section summarizes and discusses the results.

\section{A chameleon scalar field:}
The relevant action in gravity along with a chameleon field $\phi$ is given by
\begin{equation} \label{action}
A = \int\left[\frac{R}{16 \pi G} + \frac{1}{2}{\phi}_{,\mu}{\phi}^{,\mu} 
                 -V(\phi) + f(\phi) L_{m}\right]{\sqrt{-g}}~d^4x,
\end{equation}
where R is Ricci scalar, G is the Newtonian constant of gravity,$V(\phi)$ is the potential. Here $f(\phi)$ is a function of the chameleon field and determines the non-minimal coupling of the chameleon scalar field with the matter sector given by $L_{m}$. In what follows, $L_{m}$ is given by $-\rho_m$ alone. This indeed is not a standard description. However, for a collection of collisionless particles with non-relativistic speed, the rest mass of the particles completely dominates over the kinetic energy, the contribution from the pressure is negligible compared to the rest energy. The density is defined to be the contribution to that by these collisionless particles. This is thus quite a legitimate choice\cite{harko1, harko2} for a pressureless fluid, which would be relevant for the dark matter distribution.\\

By varying the action with respect to the metric tensor components, one can find the field equations. For a spatially flat FRW spacetime given by the line element

\begin{equation} \label{metric}
ds^2 = dt^2 - a^2(t) \left( dr^2 + r^2d{\theta}^2 + r^2\sin^2{\theta} 
                      d{\phi}^2 \right),
\end{equation}

 the field equations are written as

\begin{equation} \label{cons}
3\frac{{\dot{a}}^2}{a^2} = \rho_{m}f + \frac{1}{2}{\dot{\phi}}^2 + V(\phi),
\end{equation}

\begin{equation} \label{H}
2\frac{\ddot{a}}{a} + \frac{{\dot{a}}^2}{a^2} = -\frac{1}{2}{\dot{\phi}}^2 
                                                 + V(\phi),
\end{equation}
where the units are so chosen that $8 \pi G=1$. The fluid is taken in the form of pressure less dust $(p_m=0)$ consistent with a matter dominated universe and $ \rho_{m} $ is the matter density. Overhead dots denote differentiation with respect to the cosmic time $t$.\\

By varying the action with respect to the chameleon field $\phi$,  one can also find the wave equation as,
\begin{equation} \label{wave}
\ddot{\phi} + 3 H \dot{\phi} = -\frac{dV}{d\phi} - \rho_{m} \frac{df}{d \phi}.
\end{equation}

Equations (\ref{cons}), (\ref{H}) and (\ref{wave}) can be used to yield the equation
\\
\begin{equation} \label{rho}
(\rho_{m} f)\dot{~} + 3 H (\rho_{m} f) = \rho_{m} \dot{\phi} \frac{df}{d \phi},
\end{equation}
which is actually the matter conservation equation.
On integration, the equation (\ref{rho}) yields
\begin{equation} \label{rhom}
\rho_{m} = \frac{\rho_{0}}{a^3}.
\end{equation} 
All of these equations (\ref{cons}), (\ref{H}), (\ref{wave}) and (\ref{rho}) are not independent. Any one of the last two can be derived from the other three as a consequence of the Bianchi identities. We take  (\ref{cons}), (\ref{H}) and (\ref{wave}) to constitute the system of equations. There are, however, four unknowns, namely $a, \phi, V, f$. The other variable, ${\rho}_{m}$, is known in terms of $a$ via the equation (\ref{rhom}). It is intriguing to note that even with the nonminimal coupling with the scalar field, the matter energy density itself still redshifts as $\frac{1}{a^3}$ (see reference \cite{nbsdkg})\\

\section{Autonomous system:}
By introducing the following dimensionless variables  \\
\\
$x= \frac{\dot{\phi}}{\sqrt{6} H}$, $y = \frac{\sqrt{V}}{\sqrt{3} H}$, $z=\frac{\sqrt{\rho_{m} f}}{\sqrt{3} H}$, $\lambda=-\frac{1}{V}\frac{dV}{d\phi}$, $\delta=-\frac{1}{f}\frac{df}{d\phi}$, 
$\Gamma = {V \frac{d^2 V}{d \phi^2}}/{(\frac{dV}{d \phi})^2}$ and $\tau = {f \frac{d^2 f}{d \phi^2}}/{(\frac{df}{d \phi})^2}$,
\\
\\
the system of equations reduces to the following set,

\begin{equation}
x^{\prime} = -3 x + \sqrt{\frac{3}{2}} \lambda y^2 + \sqrt{\frac{3}{2}} \delta z^2 + x (3 x^2 + \frac{3}{2} z^2),
\end{equation}
\begin{equation}
y^{\prime} = - \sqrt{\frac{3}{2}} \lambda x y + y (3 x^2 + \frac{3}{2} z^2),
\end{equation}

\begin{equation}
z^{\prime} = - \frac{3}{2} z - \sqrt{\frac{3}{2}} \delta x z + z (3 x^2 + \frac{3}{2} z^2),
\end{equation}

\begin{equation}
\lambda^{\prime} = - \sqrt{6} \lambda^2 (\Gamma -1) x,
\end{equation}

\begin{equation}
\delta^{\prime} = - \sqrt{6} \delta^2 (\tau -1) x.
\end{equation}

A `prime' indicates differentiation with respect to $N = \ln a$. One can write equation (\ref{cons}) in terms of these new variables as 
\begin{equation}
x^2 + y^2 + z^2 =1.
\end{equation} 

We use this equation as a constraint equation and the system which now effectively reduces to,

\begin{equation} \label{x}
x^{\prime} = -3 x + \sqrt{\frac{3}{2}} \lambda (1-x^2-z^2) + \sqrt{\frac{3}{2}} \delta z^2 + x (3 x^2 + \frac{3}{2} z^2),
\end{equation}

\begin{equation} \label{z}
z^{\prime} = - \frac{3}{2} z - \sqrt{\frac{3}{2}} \delta x z + z (3 x^2 + \frac{3}{2} z^2),
\end{equation}

\begin{equation} \label{lambda}
\lambda^{\prime} = - \sqrt{6} \lambda^2 (\Gamma -1) x,
\end{equation}

\begin{equation} \label{delta}
\delta^{\prime} = - \sqrt{6} \delta^2 (\tau -1) x.
\end{equation}

Before getting into the actual analysis of the system, we discuss very briefly the methods that will be used in this work. Let us consider a system of linear differential equations\\
$x^{\prime} =f(x)= A x$, \\
where $x^{\prime} = \frac{dx}{dt}$, $x \equiv (x_1, x_2, ......x_n) \in \mathbb{R}^n$ and $A$ is an $n \times n$ matrix. The solutions of the system of differential equations, $A x =0$, yields the fixed points($x_0$) of the system. The stability of the fixed points can be analysed from sign of the eigen values of $A$. If all the eigen values are negative then the fixed point is a stable fixed point otherwise it is an unstable one or a saddle. For a 2D system this stability analysis can also be done from sign of the trace(Tr) and determinant(Det) of $A$. If $Tr < 0, Det > 0$, then the fixed point is stable and if $Tr >0, Det > 0$ , then the fixed point is an unstable fixed point.\\

For a system of nonlinear equations, the situation is a bit more complicated. Let us consider a system of non linear differential equations, written in the form
\begin{equation}
x^{\prime} = f(x).
\end{equation}
 Fixed points($x_0$) are the solutions of the simultaneous  equations, $f(x) = 0$. A non linear system of differential equations can be linearised near a fixed point as 
\begin{equation}
u^{\prime} = Df(x)\mid_{x_{0}} u = A u,
\end{equation}
where $u = x-x_0 $ and $A= Df(x)= (\frac{\partial f_i}{\partial x_j}), i,j=1,2,....n$. A is the Jacobian matrix of the linearised system. From the eigen values of the Jacobian matrix A at the fixed points, one can find the stability of the solutions. If all the eigen values of the Jacobian matrix at a fixed point have non vanishing real part then the fixed point is called a hyperbolic fixed point. According to Hartman-Grobman theorem, the phase portrait near a hyperbolic fixed point of a non linear system is locally equivalent to the phase portrait of the linearised system. One can use linear stability analysis to find the stability of the hyperbolic fixed points. If all the eigen values at a fixed point has negative real part, then the fixed point is a stable fixed point and if all the eigen values has positive real part , then the fixed point is an unstable fixed point. If some eigen values have negative real part and remaining eigen values have positive real part, then the
  fixed point is a saddle fixed point.\\
\\
 For non hyperbolic fixed points, one can not use linear stability analysis. In the absence of a proper analytical process, a different strategy in such cases may be adopted. In the present work, such solutions are numerically perturbed around the fixed points to check the stability in non hyperbolic cases. If the perturbed solutions asymptotically approach the fixed points, the corresponding fixed points are considered stable. This approach is quite standard in nonlinear dynamics\cite{strogatz} and  has already been utilised quite recently in a cosmological scenario\cite{nrnb}.

\section{Stability analysis of the chameleon model:}

From definition of $\Gamma$ and $\tau$, it can be easily shown that $\Gamma = 1$ corresponds to the exponential form of the potential and $\tau = 1$ corresponds to  exponential form of the coupling. For the sake of making the system of equations a bit more tractable, we classify our system into four classes as described below. \\
\\
Class I : When $\Gamma \neq 1$ and $\tau \neq 1$, i.e.,  both the potential $V(\phi)$ and the coupling $f(\phi)$ are non-exponential functions of the chameleon field.
\\
\\
Class II : When $\Gamma=1$ and $\tau \neq 1$, the potential $V$ is an exponential function of the chameleon field but the coupling $f$ is any function of the chameleon field excluding an exponential function.
\\
\\
Class III: When $\Gamma \neq 1$ and $\tau = 1$ , the potential $V$ is any function excluding an exponential function but the coupling is an exponential function of the chameleon field .
\\
\\
Class IV : When $\Gamma = 1$ and $\tau = 1$, both the potential $V$ and the coupling $f$  are exponential functions of the chameleon scalar field $\phi$.

\subsection{Class I :When both the potential and the coupling with matter are non exponential functions of the chameleon field.}

In this class, the system formed by the equations (\ref{x}),(\ref{z}), (\ref{lambda}) and (\ref{delta}), has the fixed points as listed in Table 1.\\

\begin{table}[H] 
\caption{Fixed points of Class I type models}
%\begin{ruledtabular}
\begin{tabular}{|c|c|c|c|c|c|}
\hline 
 Points & $p_1$ & $p_2$ & $p_3$ & $p_4$ & $p_5$ \\ 
\hline 
$x$ & 0 & 0 & 0 & -1 & +1 \\ 
\hline 
$z$ & 0 & +1 & -1 & 0 & 0 \\ 
\hline 
$\lambda$ & 0 & $\lambda$ & $\lambda$ & 0 & 0 \\ 
\hline 
$\delta$ & $\delta$ & 0 & 0 & 0 & 0 \\ 
\hline 
\end{tabular}
%\end{ruledtabular}
\end{table}

~\\
\\
\\
In order to study the stability of the fixed points, the eigen values of the Jacobian matrix have been found out for all the fixed points which are shown in Table 2.

\begin{table}[H] 
\caption{Eigen values of the fixed points of Class I type models}
%\begin{ruledtabular}
\begin{tabular}{|c|c|c|c|c|}
\hline 
 Points & $\mu_1$ & $\mu_2$ & $\mu_3$ & $\mu_4$ \\ 
\hline 
$p_1$ & $-3$ & $-\frac{3}{2}$ & 0 & 0 \\ 
\hline 
$p_2$ & 3 & $-\frac{3}{2}$ & 0 & 0 \\ 
\hline 
$p_3$ & 3 & $-\frac{3}{2}$ & 0 & 0 \\ 
\hline 
$p_4$ & 6 & $\frac{3}{2}$ & 0 & 0 \\ 
\hline 
$p_5$ & 6 & $\frac{3}{2}$ & 0 & 0 \\ 
\hline 
\end{tabular} 
%\end{ruledtabular}
\end{table}

~\\
\\
\linebreak
As the Jacobian matrix of the fixed points all have at least one zero eigen values, these are non hyperbolic fixed points. One can not use the  linear stability analysis in this situation. Fixed points $p_2,p_3,p_4,$ and $p_5$ has at least one positive eigen value, so these fixed points are in fact unstable but the one at $p_1$ has two negative eigen values and two zero eigen values and thus requires further investigation. It must be made clear that $p_1$ is actually an infinite set of fixed points as the value of $\delta$ is arbitrary in this case, as indicated in Table 1. We perturb the system around the fixed points $p_1$. It is not possible to draw a 4D phase plot and the 3D phase plot looks too obscure to draw conclusions from. We adopt the following strategy. We plot the projection of perturbations on $x,z,\lambda$ and $\delta$ axes separately. In figure 1, the whole $x=0$ line corresponds to the fixed point. For a small perturbation of the solution around the fixed poi
 nt $p_1$, the evolution of the solution with $N$ is studied numerically. It is evident from figure 1 that the perturbed solutions asymptotically approach $x=0$ as $N \longrightarrow \infty$. Figure 2 and 3 show that the projection of perturbations around  $z=0$ and $\lambda=0$ approaches  $z=0$ and $\lambda=0$ respectively as $N \longrightarrow \infty$.  It is interesting to note that the results given by this perturbation technique is completely consistent with the finding that $\delta$ is arbitrary. Any perturbation of the system around any value of $\delta$ renders it remaining constant at the perturbed value without any further evolution (figure 4). From this behaviour of the system near $p_1$, we can conclude that $\delta$ axis is an attractor line\cite{strogatz}. \\

Any heteroclinic orbit in the phase space of the system starts from an unstable fixed point and ends at a stable fixed point. So the universe is apt to start evolving from one of these unstable fixed points $p_2$ to $p_5$ as a result of any small perturbation and approach towards the $\delta$ axis which is an attractor line as $N \longrightarrow \infty$. The asymptotic behaviour of the fixed points helps us to understand the qualitative behaviour of the universe. \\

\begin{figure}[H]
\includegraphics[scale=0.2]{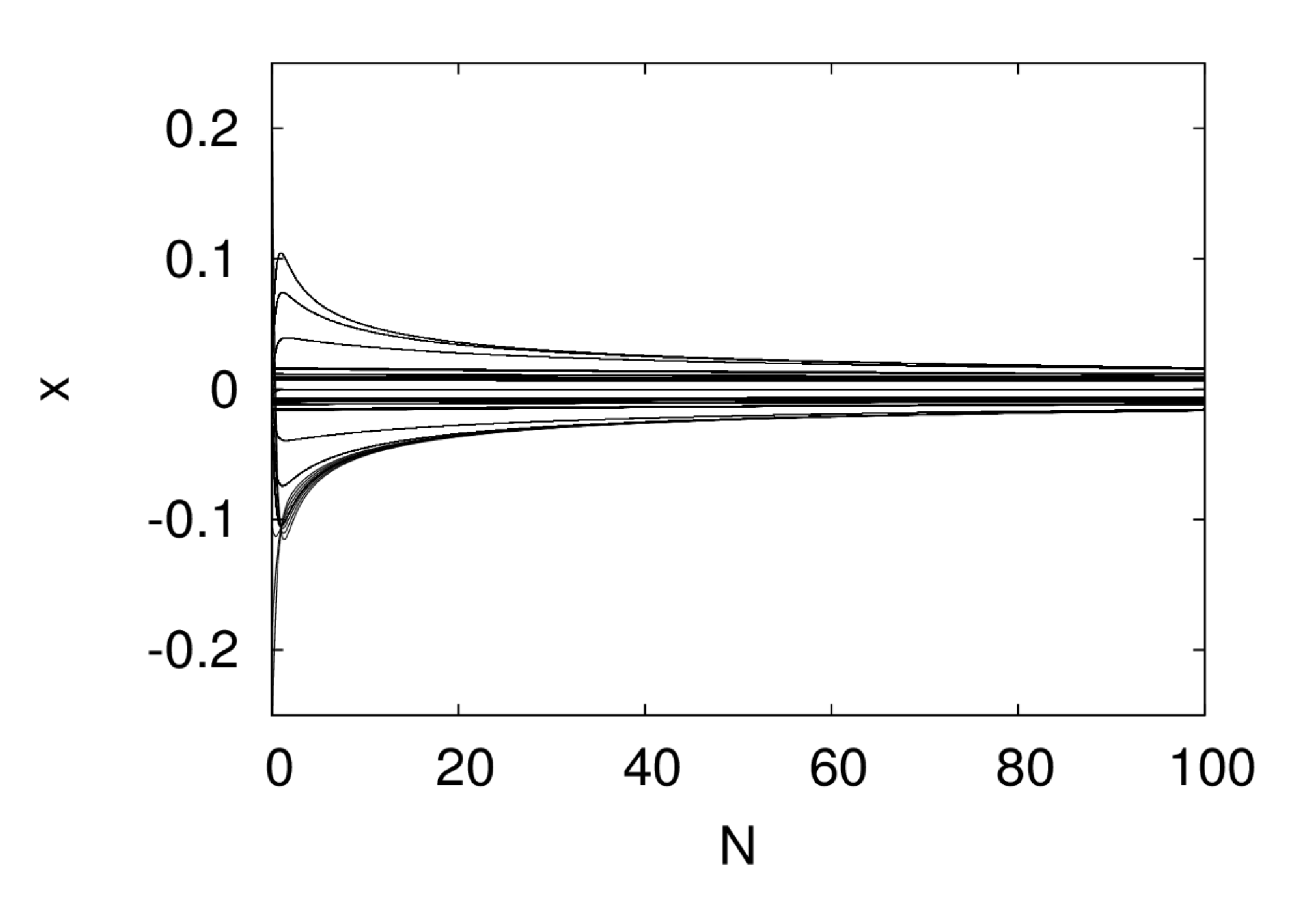} 
\caption{Plot of projections of perturbations on $x$ against N for Class I type models}
\end{figure}

\begin{figure}[H]
\includegraphics[scale=0.3]{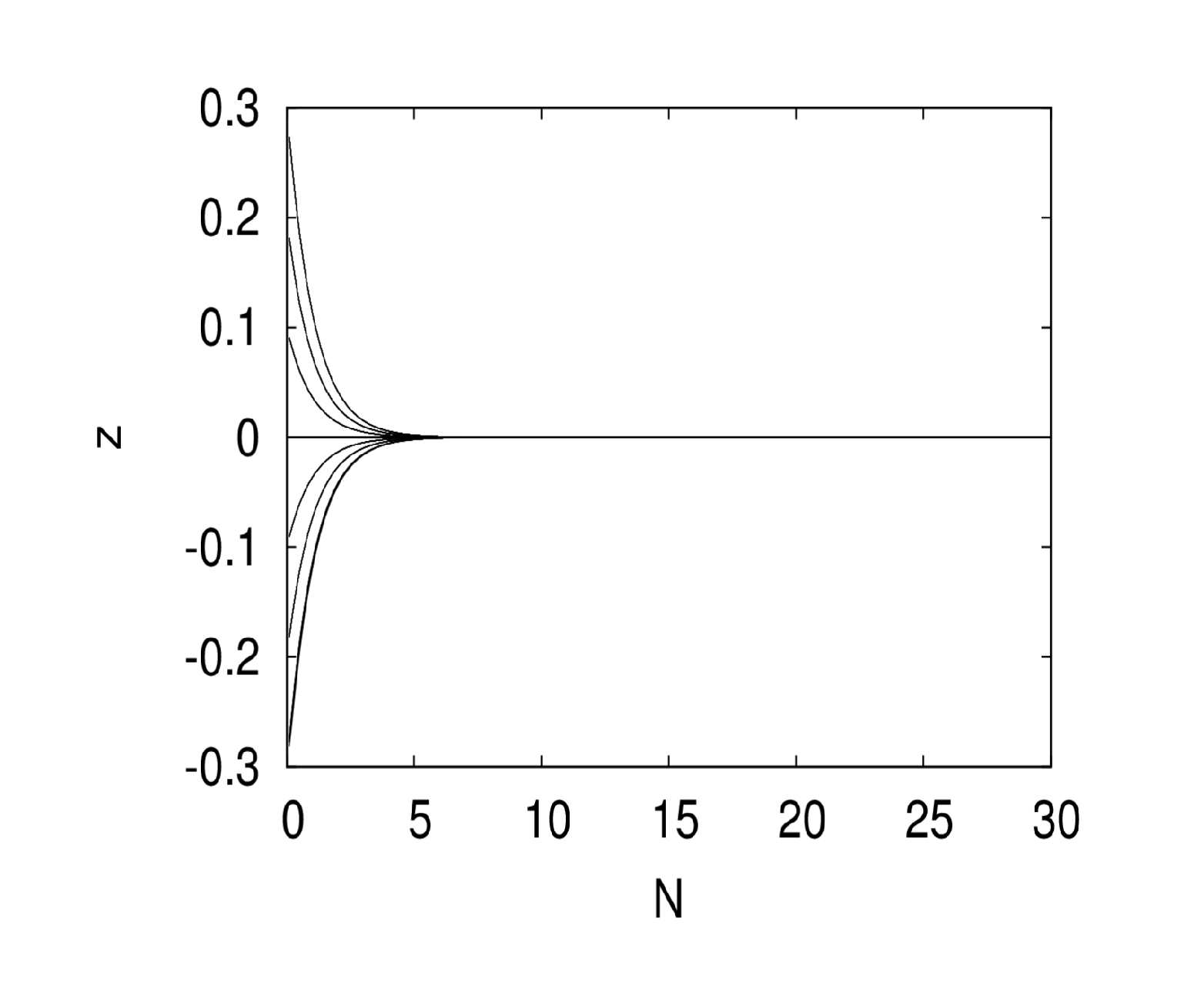}
\caption{Plot of projections of perturbations on $z$ against N for Class I type models}
\end{figure}

\begin{figure}[H]
\includegraphics[scale=0.35]{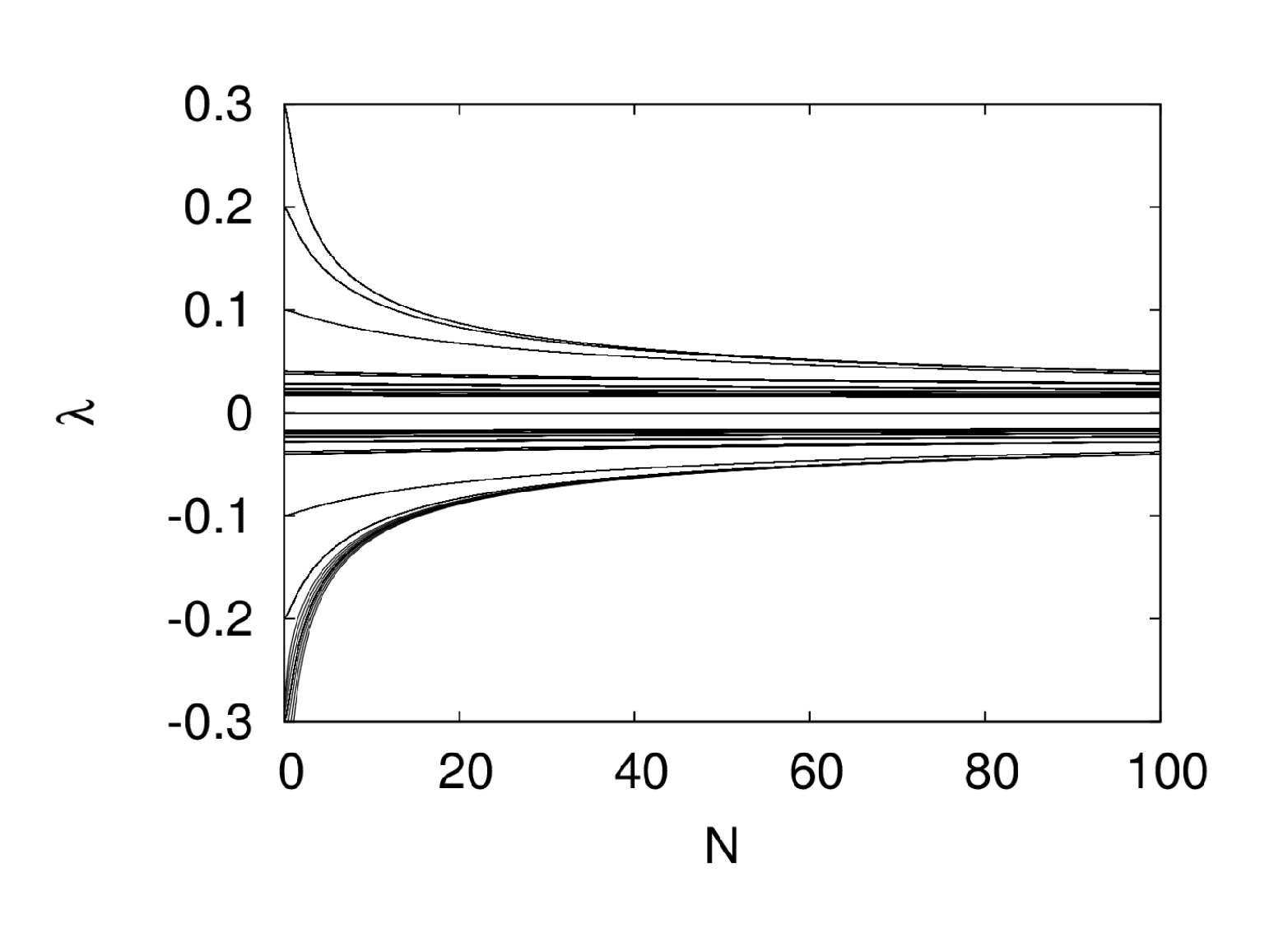}
\caption{Plot of projections of perturbations on $\lambda$ against N for Class I type models}
\end{figure}

\begin{figure}[H]
\includegraphics[scale=0.4]{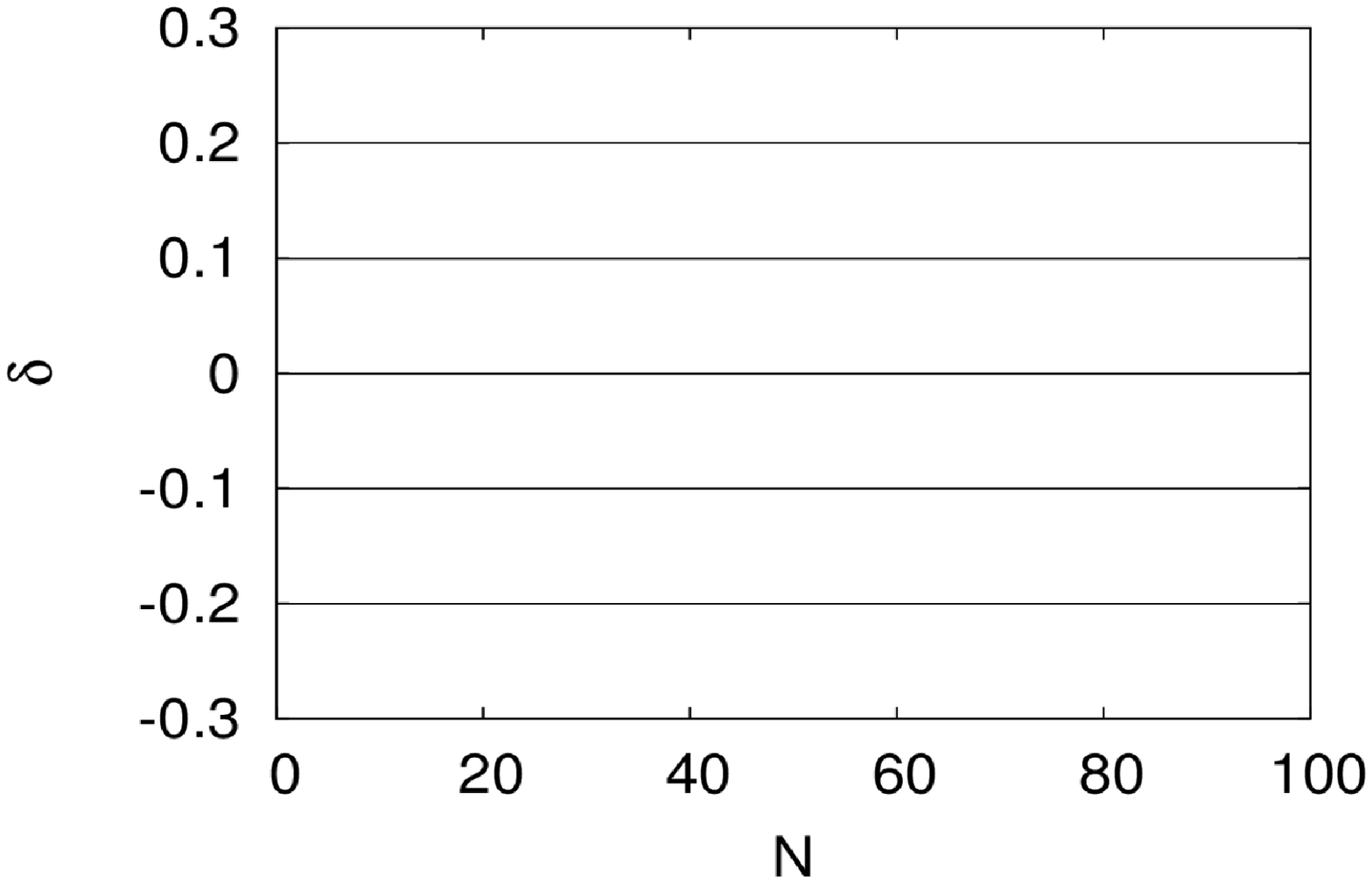}
\caption{Plot of projections of perturbations on $\delta$ against N for Class I type models}
\end{figure}

~
\linebreak
The set of fixed points $p_1$, given by $x=z=\lambda=0$ with an arbitrary $\delta$, is thus stable (an attractor) and gives the final state of the universe. The physical state of the universe in this case, as $N \longrightarrow \infty$ (i.e., $a \longrightarrow \infty$), is consistent with $\Omega_{\phi} \longrightarrow 1, ~  \Omega_{m}f \longrightarrow 0$. Here we define $\Omega_{\phi} = \frac{\frac{1}{2} \dot{\phi}^2 + V }{3 H^2}$ and $\Omega_m = \frac{\rho_m}{3 H^2}$.\\

It deserves mention that as $x=0$, the contribution from the kinetic part of the chameleon goes to zero and this complete domination of the chameleon field is entirely given by the potential part. From equation (\ref{H}), one can see that the universe settles to an accelerated phase with $q\longrightarrow -1$.\\

The unstable fixed points $p_{2} $ and $p_3 $ indicate states with $\rho_{m}\longrightarrow \infty$ and $\Omega_{\phi}\longrightarrow 0$. This is consistent with the physical requirement of the early phase of the evolution. However the fixed point $p_3$ will henceforth be disregarded as unphysical, as for a positive $H$, negative values of $z$ do not represent any physical state.\\
  
Like $p_2$, the other two fixed points $p_4$ and $p_5$ are also unstable fixed points and describe physical states with the relative magnitude of $\Omega_{m}f$ and $\Omega_{\phi}$ being reversed. However, it also deserves mention that  for both $p_4$ and $p_5$, $p_{\phi}$ remains positive (equal to $\rho_{\phi}$) and thus the scalar field does not really play the role of a dark energy. It deserves mention that $x$ and $\lambda$ as given by figures 1 and 3 might appear to be leading to a non-zero constant value. But the ordinates are not really parallel to the horizontal axis, they do vary, but only too slowly. It is checked that they indeed approach zero for a very large value of $N$.

\subsection{Class II: When the potential is an exponential function of the chameleon field but the coupling with matter is any function of the chameleon field except an exponential one:}

In this class $\lambda$ is a constant and equations (\ref{x}), (\ref{z}) and (\ref{delta}) form the system of equations. The fixed points of the system are given in Table 3 and the eigenvlaues of the Jacobian matrix at the fixed points are given in Table 4.\\

\begin{table}[H] 
\caption{Fixed points of Class II type models}
%\begin{ruledtabular}
\begin{tabular}{|c|c|c|c|c|c|c|c|c|}
\hline 
Points & $q_1$ & $q_2$ & $q_3$ & $q_4$ & $q_5$ & $q_6$ & $q_7$ & $q_8$\\ 
\hline 
x & 0 & 0 & 0 & +1 & -1 & $\frac{\lambda}{\sqrt{6}}$ & $\sqrt{\frac{3}{2 \lambda^2}}$ & $\sqrt{\frac{3}{2 \lambda^2}}$\\ 
\hline 
z & 0 & +1 & -1 & 0 & 0 & 0 & $ \sqrt{1-\frac{3}{\lambda^2}} $ & $ -\sqrt{1-\frac{3}{\lambda^2}} $ \\ 
\hline 
$\delta$ & $\delta$ & 0 & 0 & 0 & 0 & 0 & 0 & 0 \\ 
\hline 
\end{tabular} 
%\end{ruledtabular}
\end{table}
~\\

\begin{table}[H] 
\caption{Eigen values of the fixed points of Class II type models}
%\begin{ruledtabular}
\begin{tabular}{|c|c|c|c|}
\hline 
{Points}  & $\mu_1$  &  $\mu_2$ & $\mu_3$ \\ 
\hline 
$q_1$ & $-3$ & $- \frac{3}{2}$ & 0 \\ 
\hline 
$q_2$ & 3 & $- \frac{3}{2}$ & 0 \\ 
\hline 
$q_3$ & 3 & $- \frac{3}{2}$ & 0 \\ 
\hline 
$q_4$ & 3 & $6 - \sqrt{6} \lambda$ &  0 \\ 
\hline 
$q_5$ & 3 & $6 + \sqrt{6} \lambda$ &  0 \\ 
\hline 
$q_6$ & $-3 + \frac{\lambda^2}{2}$ &  $- \frac{3}{2} + \frac{\lambda^2}{2}$ & 0 \\ 
\hline 
$q_7$ and $q_8$ & $\mu_{27}$ & $\mu_{37}$ & 0 \\ 
\hline 
\end{tabular} 
%\end{ruledtabular}
\end{table}

~\\
Here  the symbols, $\mu_{27}= \frac{3}{4} (-1-\frac{1}{\lambda} \sqrt{24- 7 \lambda^2})$ and $\mu_{37}= \frac{3}{4} (-1+\frac{1}{\lambda} \sqrt{24- 7 \lambda^2})$, are used for brevity.\\

 Existence of the fixed point at $q_1$ demands $\lambda =0$, indicating that $V$ is a constant in this case. This could have a bearing on the chameleon mechanism. This is a non isolated fixed point and has one zero eigen value at each point, defined by the value of $\delta$, of the equilibrium set. These type of fixed points are called normally hyperbolic fixed points (\cite{coley}, \cite{ellis}). The stability of a normally hyperbolic fixed point depends on the signatures of the remaining eigenvalues. If sign of the remaining eigenvalues are negative then the fixed point is a stable fixed point otherwise the fixed point is an unstable one. For $q_1$, the remaining eigen values are $-3$ and $-\frac{3}{2}$, so $q_1$ is a stable fixed point.\\
 
Other fixed points are  non hyperbolic and not even normally hyperbolic. The fixed point $q_6$ can have one zero eigen value and two negative eigen values if $\lambda^2 < 3$ whereas $q_7$ can also have one zero eigen value with two negative eigen values if $\frac{24}{7} \leq \lambda^2 \leq 3$. So in the corresponding  parameter range $q_6$ and $q_7$ need further investigation as in case of Class I type of models. It is found that for a small perturbation about these fixed points ($q_6$, $q_7$ and $q_8$), the solutions do not approach towards the fixed points as $N \longrightarrow \infty$, so these fixed points are unstable. The fixed points $q_2, q_4$ and $q_5$ are also unstable fixed points as they have at least one positive eigen value. So $q_2$ to $q_8$ could be the beginning of the evolution and $q_1$  may be the possible ultimate fate of the universe. It also deserves mention that for $\lambda =0$, $q_6$ is in fact a subset of $q_1$. It is interesting to note that for $\lambda = 0$, the fixed point $q_7$ is realized only for $x \longrightarrow \infty$.  The fixed points $q_3$ and $q_8$ are not analyzed on the plea that for a negative $z$, the state is not physical as it corresponds to a negative nalue of $H$.\\

Amongst the fixed points, potentially the beginning of the universe (i.e., unstable fixed points), $q_2$ and $q_3$ has the possibility of $\Omega_{m}f =1$ and $\Omega_{\phi} =0$ which is a desired situation. The only stable fixed point, $q_1$, yields the possibility of a final fate where $\Omega_{m} f =0$ and $\Omega_{\phi} =1$ with $q\longrightarrow -1$.

\subsection{Class III :When the coupling with matter is an exponential function of the chameleon field but the potential is any function of the same except an exponential function:}
When the coupling of the chameleon with the matter field is an exponential function of $\phi$, one has $\delta =$ constant. The system of equations is formed by equations (\ref{x}), (\ref{z}) and (\ref{lambda}). Fixed points of the system are given in Table 5 and the eigen values of the Jacobian matrix at the fixed points are given in Table 6.\\

\begin{table}[H] 
\caption{Fixed points of Class III type models}
%\begin{ruledtabular}
\begin{tabular}{|c|c|c|c|c|c|c|c|}
\hline 
Points & $m_1$ & $m_2$ & $m_3$ & $m_4$ & $m_5$ & $m_6$ &$m_7$ \\ 
\hline 
x & 0 & 0 & 0 & 1 & -1 & $\sqrt{\frac{2}{3}} \delta$ & $\sqrt{\frac{2}{3}} \delta$ \\ 
\hline 
z & 0 & $+1$ & $-1$ & 0 & 0 & $\sqrt{1-\frac{2 \delta^2}{3}}$ & $-\sqrt{1-\frac{2 \delta^2}{3}}$ \\ 
\hline 
$\lambda$ & 0 & $\lambda$ & $\lambda$ & 0 & 0 & 0 & 0 \\ 
\hline 
\end{tabular} 
%\end{ruledtabular}
\end{table}

\begin{table}[H] 
\caption{Eigen values of the fixed points of Class III type models}
%\begin{ruledtabular}
\begin{tabular}{|c|c|c|c|}
\hline 
Points & $\mu_1$ & $\mu_2$ & $\mu_3$ \\ 
\hline 
$m_1$ & $-3$ & $-\frac{3}{2}$ & 0 \\ 
\hline 
$m_2$ & 3 & $-\frac{3}{2}$ & 0 \\ 
\hline 
$m_3$ & 3 & $-\frac{3}{2}$ & 0 \\ 
\hline 
$m_4$ & 6 & $\frac{3}{2} - \sqrt{\frac{3}{2}}\delta$ & 0 \\ 
\hline 
$m_5$ & 6 & $\frac{3}{2} + \sqrt{\frac{3}{2}}\delta$ & 0 \\ 
\hline 
$m_6$ & $\frac{1}{2} (-3+2 \delta^2)$ & $(3+2 \delta^2)$ & 0  \\ 
\hline 
$m_7$ & $\frac{1}{2} (-3+2 \delta^2)$ & $(3+2 \delta^2)$  & 0\\ 
\hline 
\end{tabular} 
%\end{ruledtabular}
\end{table}
%\end{center}
~\\
The existence of fixed point $m_2$ requires the condition that $\delta=0$ indicating that $f$ is a constant! This would require that the nonminimal coupling between the chameleon field and matter becomes trivial and certainly there is a breakdown of the chameleon mechanism. Unphysical states (negative values of $z$) given by the fixed points $m_3$ and $m_7$ are not discussed.\\

 Fixed point $m_1$ has one zero eigen value and two negative eigen values. It is a non hyperbolic fixed point and thus one can not use the linear stability analysis in this case. So we perturbed the system from the fixed point and numerically solved the system of equations for  each perturbations. Plots of the projections of the perturbations on $x,z$ and $\lambda$ against $N$ in figures (5), (6) and (7) show quite convincingly that for large $N$, the solutions approach the fixed point which is thus an attractor. So $m_1$ is apt to represent the final state of the evolution.\\
 
  Other fixed points are unstable. In the beginning, the universe could be at any of these unstable fixed points and with a small perturbation from these unstable fixed points might start evolving towards the attractor $m_1$ which can be the possible final state of the universe for which $\Omega_{m}f \longrightarrow 0$ and $\Omega_{\phi} \longrightarrow 1$ are evident possibilities. The final stage is completely dominated by the potential $V$ which is a constant, effectively giving a cosmological constant, with $q\longrightarrow -1$.\\
  
   Amongst the unstable fixed point, $m_2$ is definitely favoured as they indicate $\Omega_{m}f \longrightarrow 1$ and $\Omega_{\phi}\longrightarrow 0$ for  the initial epoch. For the unstable fixed points $m_4$ and $m_5$, although $\Omega_{\phi}\longrightarrow 1$, the chameleon cannot act as a dark energy as $p_{\phi}$ is not negative. The fixed point  $m_6$ will indicate a matter dominated early epoch if $\delta \leq \frac{\sqrt{3}}{2}$ and could well be amongst the viable options. Like Case I, in this case also, $x$ and $\lambda$ actually approach zero which had been revealed on an  extension of the horizontal axis, roughly upto 1000.

\begin{figure}[H]
\includegraphics[scale=0.6]{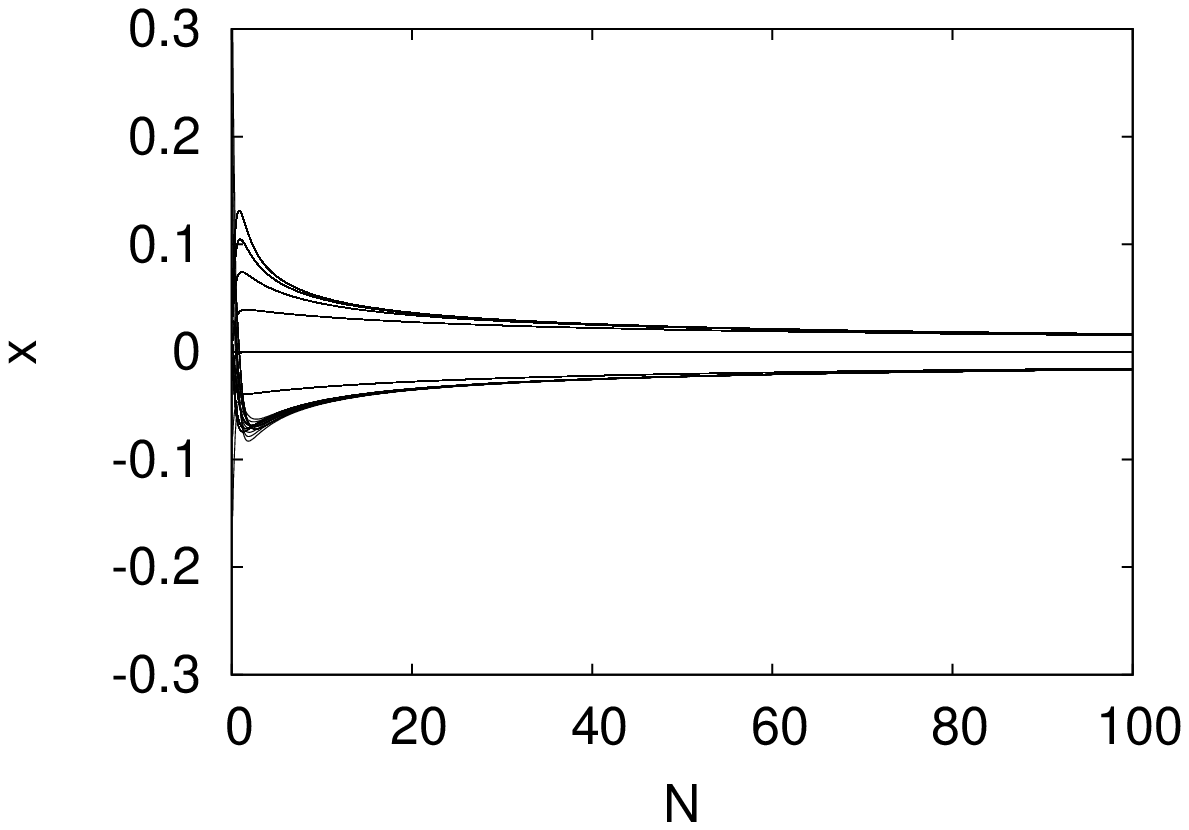}
\caption{Plot of projections on $x$ vs $N$ for Class III type of models}
\end{figure}

\begin{figure}[H]
\includegraphics[scale=0.6]{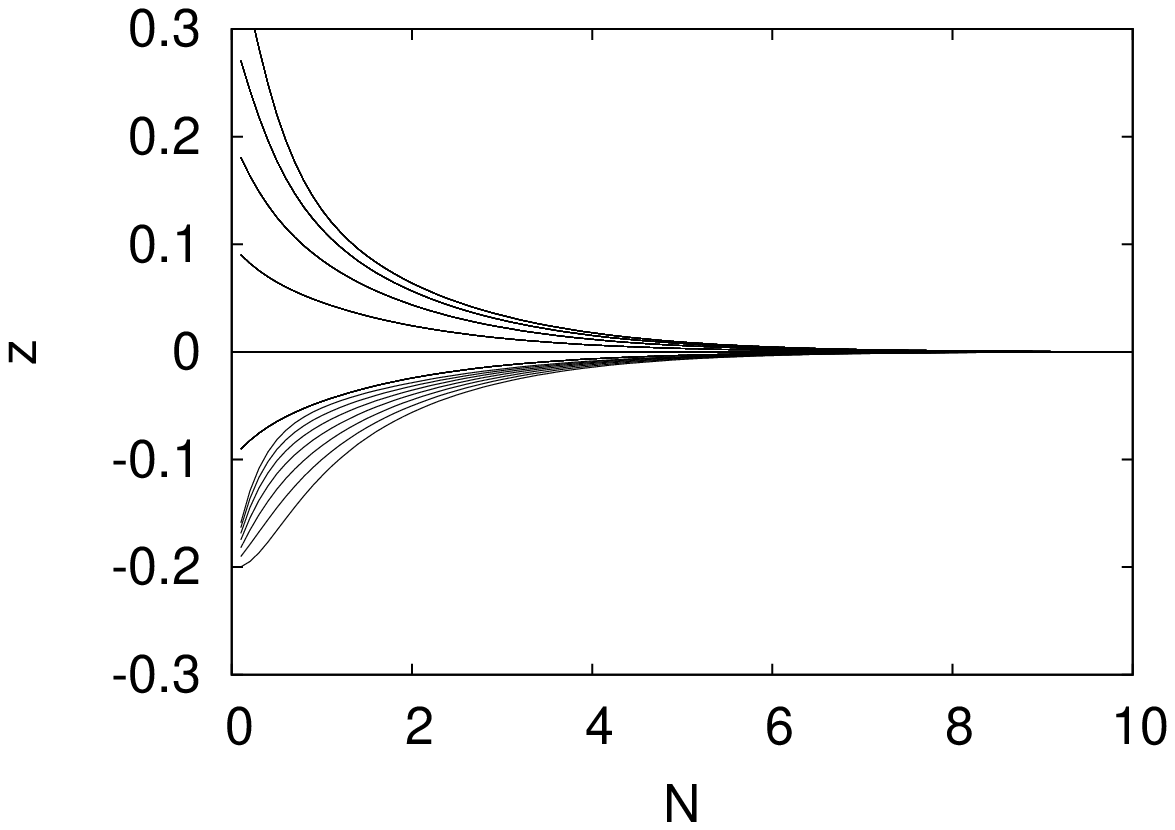}
\caption{Plot of projections on $z$ vs $N$ for Class III type of models}
\end{figure}

\begin{figure}[H]
\includegraphics[scale=0.6]{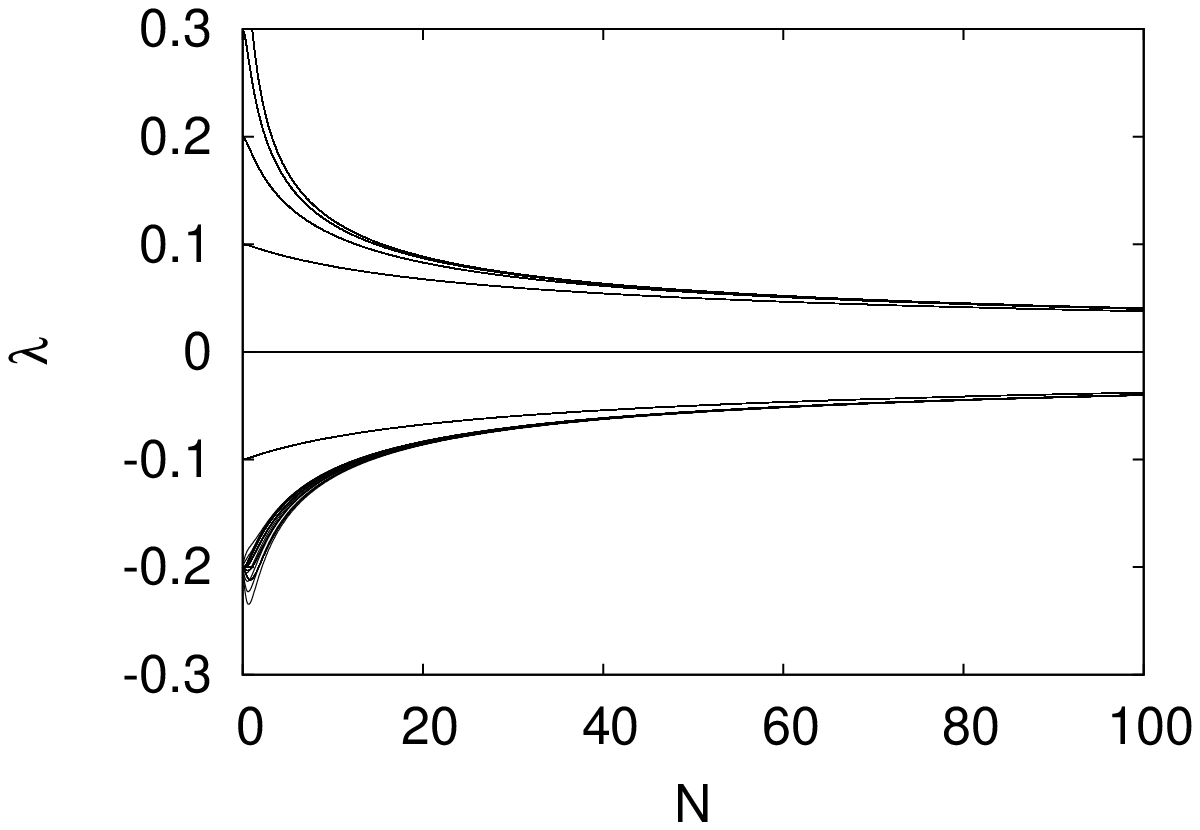}
\caption{Plot of projections on $\lambda$ vs $N$ for Class III type of models}
\end{figure}

\subsection{Class IV :When the potential and the coupling with matter both are exponential functions of the chameleon field :}
~\\
In this class of the model both $\lambda$ and $\delta$ are constants. Equations (\ref{x}) and (\ref{z}) constitute the system of equations for the effectively two-dimensional problem. The fixed points of the system are given in Table 7.\\

%\squeezetable
\begin{table}[H] 
\caption{Fixed points of Class IV type models}
%\begin{ruledtabular}
\begin{tabular}{|c|c|c|c|c|c|c|c|}
\hline 
Points & $n_1$ & $n_2$ & $n_3$ & $n_4$ & $n_5$ & $n_6$ & $n_7$ \\ 
\hline 
x & -1 & 1 & $\frac{\lambda}{\sqrt{6}}$ & $\sqrt{\frac{2}{3}} \delta$ & $\sqrt{\frac{2}{3}} \delta$ & $-\sqrt{\frac{3}{2}} \frac{1}{\delta - \lambda}$ & $-\sqrt{\frac{3}{2}} \frac{1}{\delta - \lambda}$ \\ 
\hline 
z & 0 & 0 & 0 & $-\sqrt{1- \frac{2 \delta^2}{3}}$ & $\sqrt{1- \frac{2 \delta^2}{3}}$ & $- \frac{\sqrt{-3- \delta \lambda + \lambda^2}}{\delta - \lambda}$ & $ \frac{\sqrt{-3- \delta \lambda + \lambda^2}}{\delta - \lambda}$ \\ 
\hline 
\end{tabular} 
%\end{ruledtabular}
\end{table}

~\\
\linebreak
This is a 2D system, so the stability of the fixed points can be checked by both the methods, like from sign of eigen values of the Jacobian matrix or from trace and determinant of the Jacobian matrix at the fixed point. Depending on the degree of simplicity they offer, one of these two methods are used for different fixed points. There are seven fixed points, $n_1$ to $n_7$. We analyse the stability of the fixed points $n_1$ to $n_5$ by the signature of the  eigenvalues of corresponding Jacobian matrix (see Table 8). For $n_6$ and $n_7$, the trace and determinant of the Jacobian matrix are looked at. For the present case when both $V$ and $f$ are exponential functions of $\phi$, both of $\lambda$ and $\delta$ are constants and the fixed points are found in terms of these two constants of the theory. The stability criteria crucially depend on the values of $\lambda$ and $\delta$. Depending on the values of these two constants, there are special special cases. One of them, namely $n_3$ with $\lambda =0$, which results in $x=0, z=0, \lambda =0$, has an interesting feature that it actually coincides with $p_1, q_1$ and $m_1$. This fixed point will be taken up in the last section.\\

%\squeezetable
\begin{table}[H] 
\caption{Eigen values of the fixed points of Class IV type models}
%\begin{ruledtabular}
\begin{tabular}{|c|c|c|c|}
\hline 
Points &  $\mu_1$ & $\mu_2$ & Conditions of stability \\ 
\hline 
$n_1$ & $\frac{1}{2} (3 + \sqrt{6} \delta)$ & $6+\sqrt{6} \lambda$ & $\lambda<-\sqrt{6}, \delta < - \sqrt{\frac{3}{2}}$ \\ 
\hline 
$n_2$ & $\frac{1}{2} (3 - \sqrt{6} \delta)$ & $6-\sqrt{6} \lambda$ & $\lambda>\sqrt{6}, \delta > \sqrt{\frac{3}{2}}$ \\ 
\hline 
$n_3$ & $\frac{1}{2}(-6+\lambda^2)$ & $\frac{1}{2}(-3-\lambda \delta+\lambda^2)$ & $\lambda^2 < 6, \delta> \frac{\lambda^2-3}{\lambda}$ \\ 
\hline 
$n_4$ & $-\frac{3}{2} + \delta^2$ & $3+ 2 \delta^2 - 2 \delta \lambda$ & $\delta^2 < \frac{3}{2}, \lambda > \frac{3+ 2 \delta^2}{2 \delta}$ \\ 
\hline 
$n_5$ & $-\frac{3}{2} + \delta^2$ & $3+ 2 \delta^2 - 2 \delta \lambda$ & $\delta^2 < \frac{3}{2}, \lambda > \frac{3+ 2 \delta^2}{2 \delta}$ \\ 
\hline 
\end{tabular} 
%\end{ruledtabular}
\end{table}

Trace and determinant of the Jacobian matrix at $n_6$ and $n_7$ are $TrA = \frac{-6 \delta + 3 \lambda}{2(\delta - \lambda)}$ and $DetA= \frac{-3(3+2\delta (\delta - \lambda))(3 + (\delta - \lambda) \lambda}{2(\delta - \lambda)^2}$ respectively. Conditions for stability of $n_6$ and $n_7$ are,\\

1. $\lambda < -\sqrt{6} , \delta > \frac{\lambda}{2} + \frac{1}{2} \sqrt{\lambda^2 - 6}$.\\
\linebreak
2. $\lambda = - \sqrt{6}, \delta > - \sqrt{\frac{3}{2}}$.\\
\linebreak
3. $-\sqrt{6} < \lambda < 0, \delta > \frac{-3 + \lambda^2}{\lambda}$.\\
\linebreak
4. $0< \lambda < \sqrt{6}, \delta < \frac{-3 + \lambda^2}{\lambda}$.\\
\linebreak
5. $\lambda = \sqrt{6}, \delta < \sqrt{\frac{3}{2}}$.\\
\linebreak
6.$\lambda> \sqrt{6}, \delta < \frac{\lambda}{2} - \frac{1}{2} \sqrt{ 6- \lambda^2} $.\\
 
 The fixed point $n_4$ is irrelevant as it requires a negative $z$ and thus a negative $H$ and $n_6$ is relevant only in cases where $\delta - \lambda$ is negative for the same reason. In what follows, an example of an unstable and that of a stable fixed point of this class are given. We choose $\Omega_{\phi}=0.73$ and the deceleration parameter $q=0.53$ \cite{giostri} in order to pick up some relevant values of the model parameters $\lambda$ and $\delta$. With this value of $\Omega_{\phi}$, it is easy to see from the definitions of $x$ and $y$ that $x_0^{2} + y_0^{2} = 0.73$ and thus from the constraint $x^{2}+y^{2}+z^{2} = 1$, one has $ z_0^{2} = 0.27$, i.e., $z=\pm 0.51$. From the field equations (\ref{cons}) and (\ref{H}), one can write 
\begin{equation}
 q=3x^{2}+\frac{3}{2}z^{2}-1,
\end{equation}
which yields $x=\pm 0.346$. Now from the Table 7, Table 8 and list of the stability conditions of $n_6$ and $n_7$, one can check that for $\lambda = 1.8$ and $\delta = -0.1$, one has $n_1$ as an unstable fixed point and $n_6$ as a stable one. So the universe is apt to start its evolution from $n_1$ and settle into a final configuration at $n_6$. The behaviour of $x, z$ against $N$ and that of the cosmological parameters $q$, $\Omega_m$, $\Omega_{\phi}$ and $\gamma_{\phi}$ against $N$ are given in figures 8 and 9 respectively. It is easy to check that the universe starts with a deceleration ($q=2$) but with $\Omega_{\phi}=1$ and $\Omega_m =0$ and settles down to final phase of decelerated expansion with $q=0.42$ and again the same state of  $\Omega_{\phi}=1$ and $\Omega_m f =0$. There is an accelerated expansion in between, around the present stage of evolution (see figure 9).  \\

As this case effectively reduces to a 2-dimensional problem, one can draw the phase plot in $x$ and $z$, for the given values of $\lambda$ and $\delta$, 1.8 and -0.1 respectively. The plot is shown in figure 10. In figure 11, we zoom the plot around $q_6$ in order to understand the stability a bit more clearly.

\begin{figure}[H]
\includegraphics[scale=0.3]{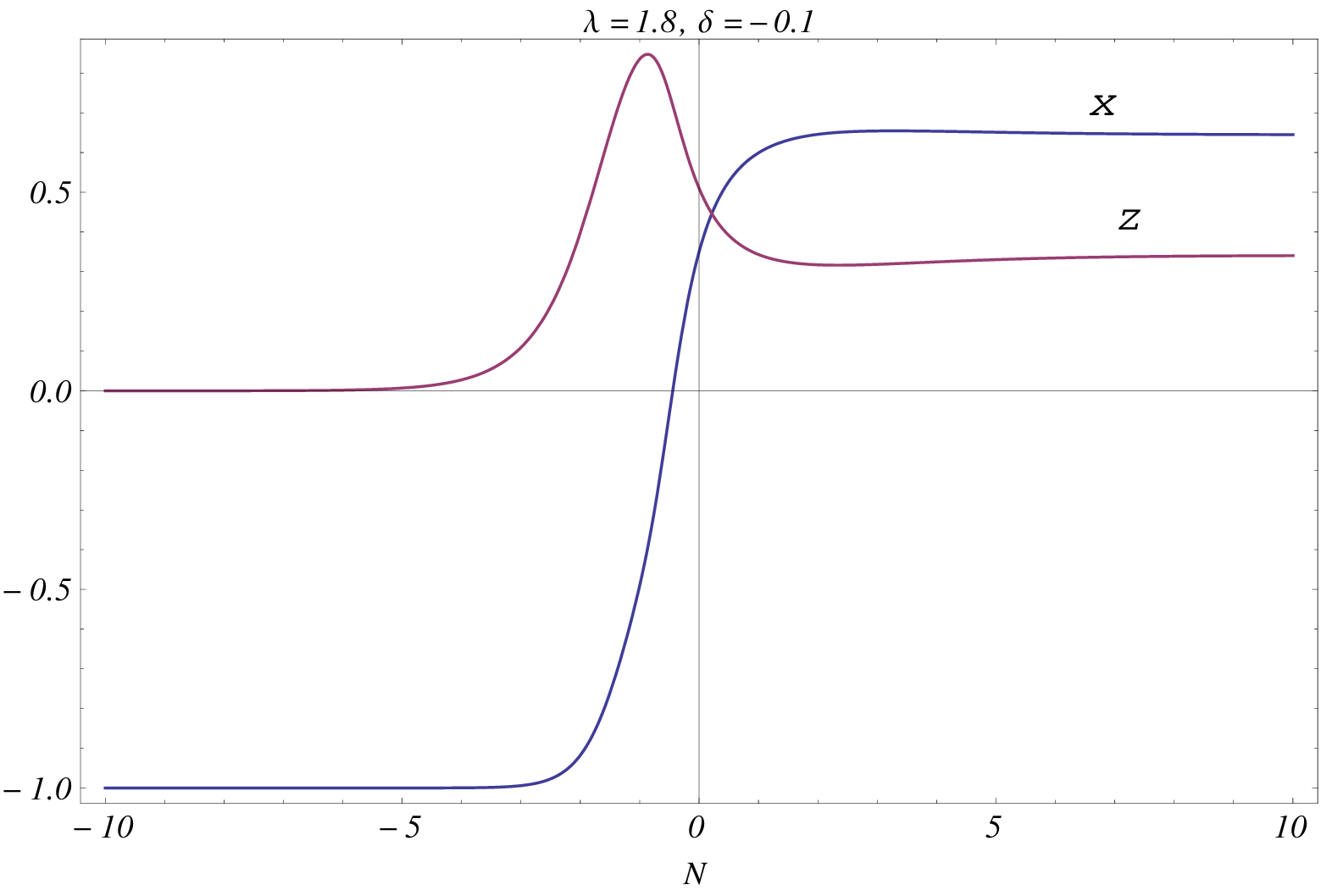}
\caption{Plot of x and z against N for Class IV models}
\end{figure}

\begin{figure}[H]
\includegraphics[scale=0.3]{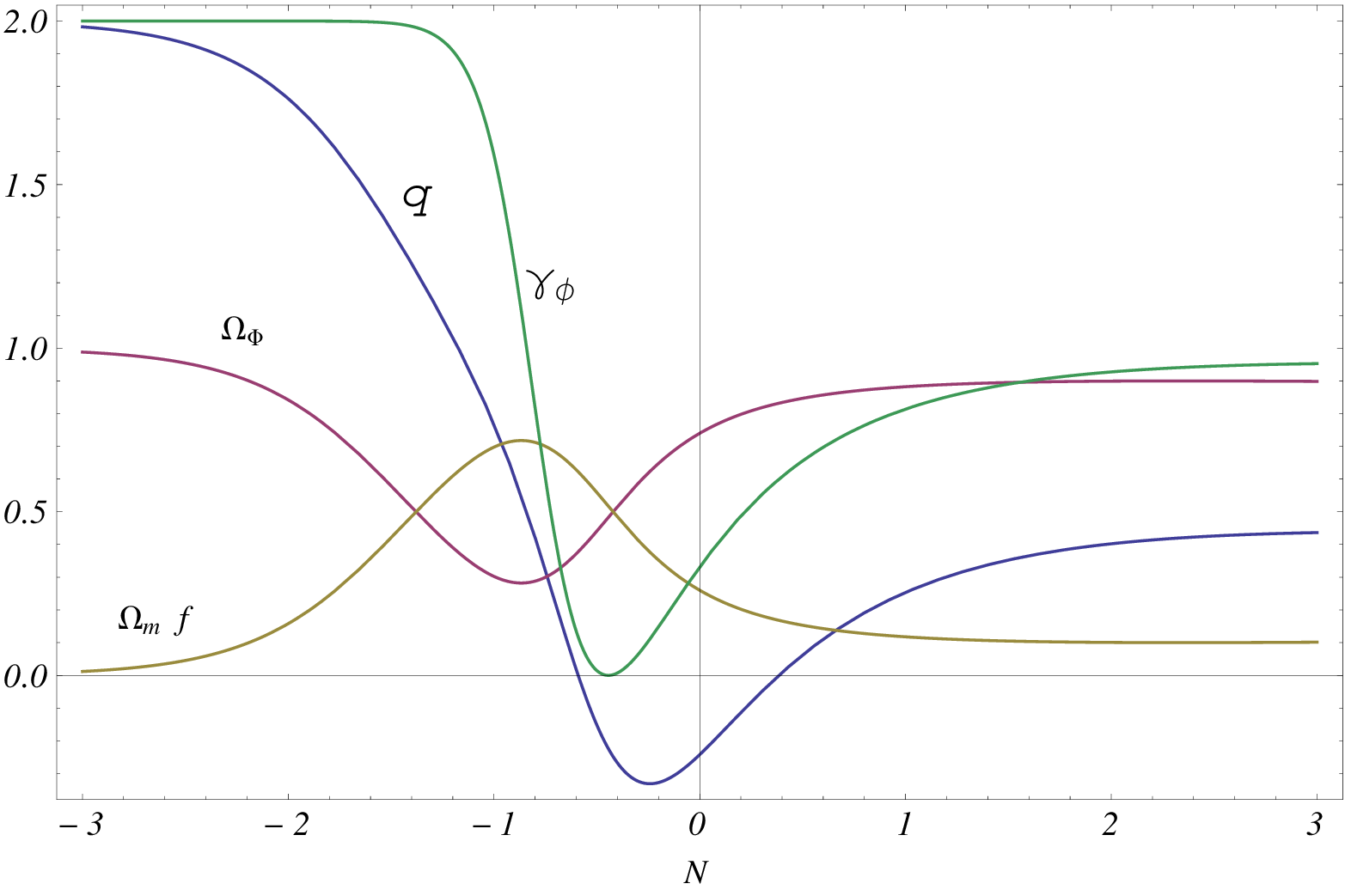}
\caption{Plot of physical parameters q, $\gamma_\phi, \Omega_{\phi} $ and $\Omega_{\phi} f$ against N for Class IV models for $\lambda = 1.8$ and $\delta = -0.1$}
\end{figure}

\begin{figure}[H]
\includegraphics[scale=0.7]{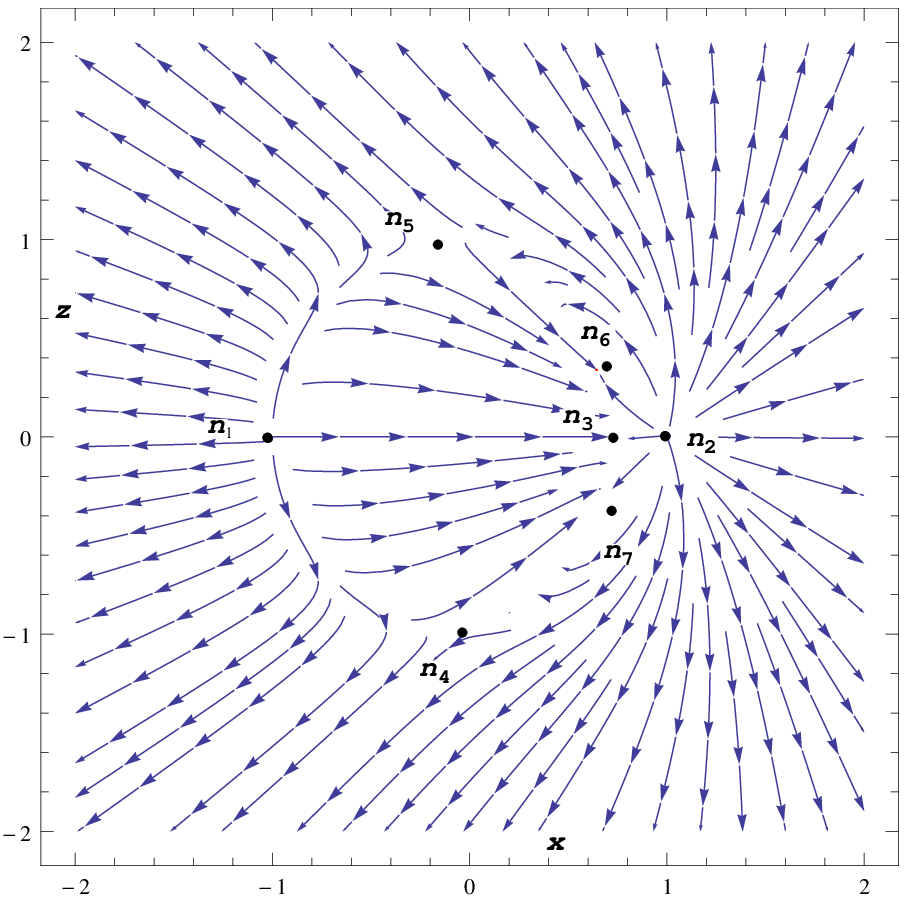}
\caption{Phase plot of the Class IV type models}
\end{figure}

\begin{figure}[H]
\includegraphics[scale=0.7]{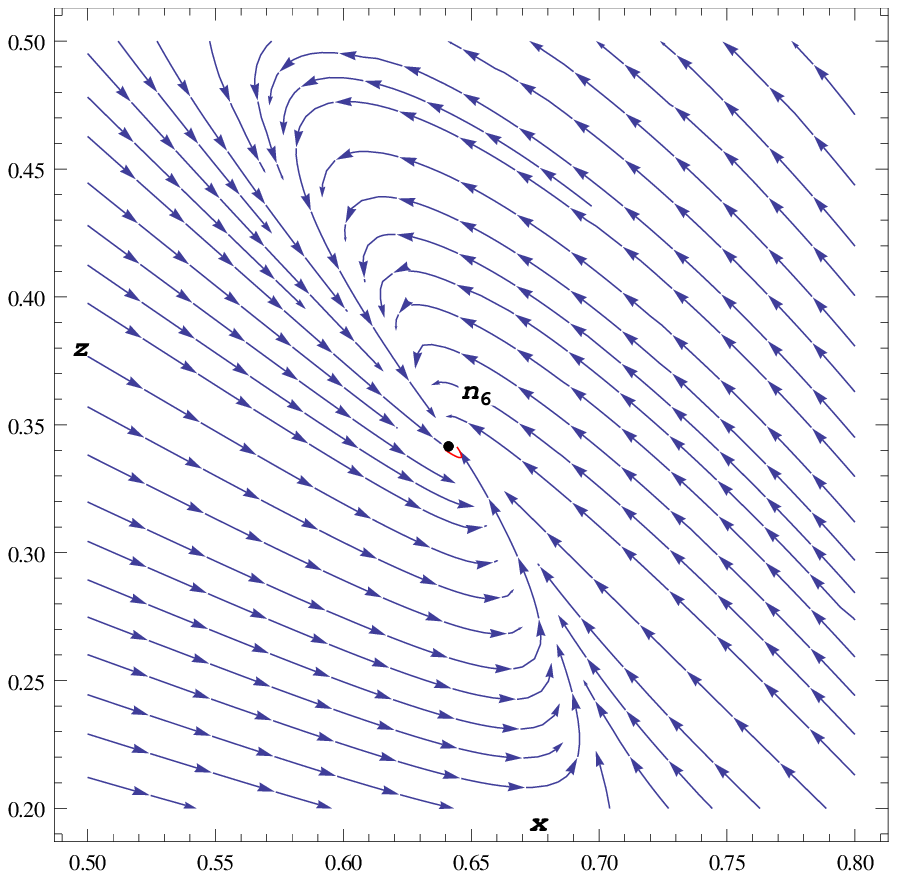}
\caption{Phase plot of the Class IV type models near $n_6$}
\end{figure}

\section{Chameleon mechanism and acceleration of the universe}

Khoury discussed the restriction of the mass of the chamaleon field vis-a-vis the coupling of the chameleon field with matter considering the compatibility of adaptation of the chameleon with the ambience and the requirement of the laboratory based experiments\cite{khoury}. The example taken up is the one for which $ V(\phi) = \frac{M^{4+n}}{\phi^n}$ and $f(\phi)= \xi \frac{\phi}{M_{pl}} \rho$. Here $M$ and $n$ are constants and $\xi$ determines the strength of the coupling $f$. The allowed band in the $m_\phi$ vs $\xi$ plot was worked out. \\

The example is clearly in the Class I of of the present work where both of $V$ and $f$ are non-exponential functions. In this case $\Gamma = 1 + \frac{1}{n^2}$ and $\tau = 0$, both are independent of $\xi$. As an illustration, we take up the case for $n=4$. With the boundary values of $x$, $z$ chosen from observational constraints as in section 4.4, one can plot the variables $x, z, \lambda$ and $\delta$ (figure12). The plot of the deceleration parameter $q$ against $N$ (figure 13) shows that the universe has a smooth transition from a decelerated to an accelerated expansion approximately at $N = -0.55$ which corresponds to a redshift of 0.74, which perfectly matches the observation\cite{farooq}. So we find one example where this transition happens independent of the coupling of the scalar field with matter and hence posing no threat to the allowed band between $\xi$ and the mass of the scalar field.  It also deserves mention that the system has starts evolving from an unstable fixed point $(1,0,0,0)$ and approaches to a stable fixed point $(0,0,0,\delta)$, which also agrees very well with the present dynamical systems analysis of ClassI models.

\begin{figure}[H]
   \begin{center}
 \includegraphics[scale=0.6]{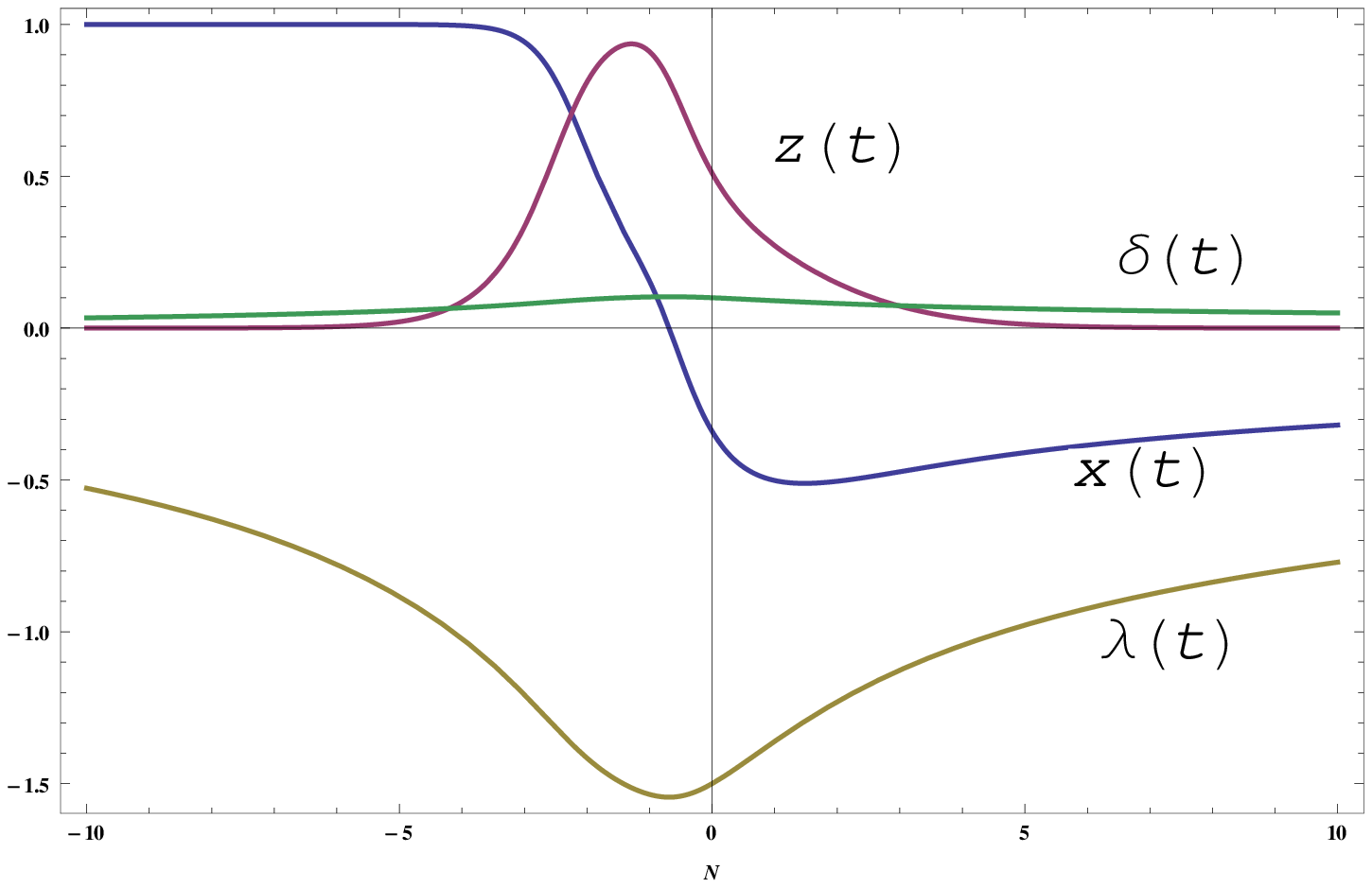}
 % sol1.eps: 0x0 pixel, 300dpi, 0.00x0.00 cm, bb=0 0 407 263
 \caption{Plot of $x,z, \lambda$ and $\delta$ against $N$ for $\lambda_0 = -1.5$ and $\delta_0 = 0.1$.}
\end{center}
 \end{figure}

 \begin{figure}[H]
   \begin{center}
 \includegraphics[scale=0.6]{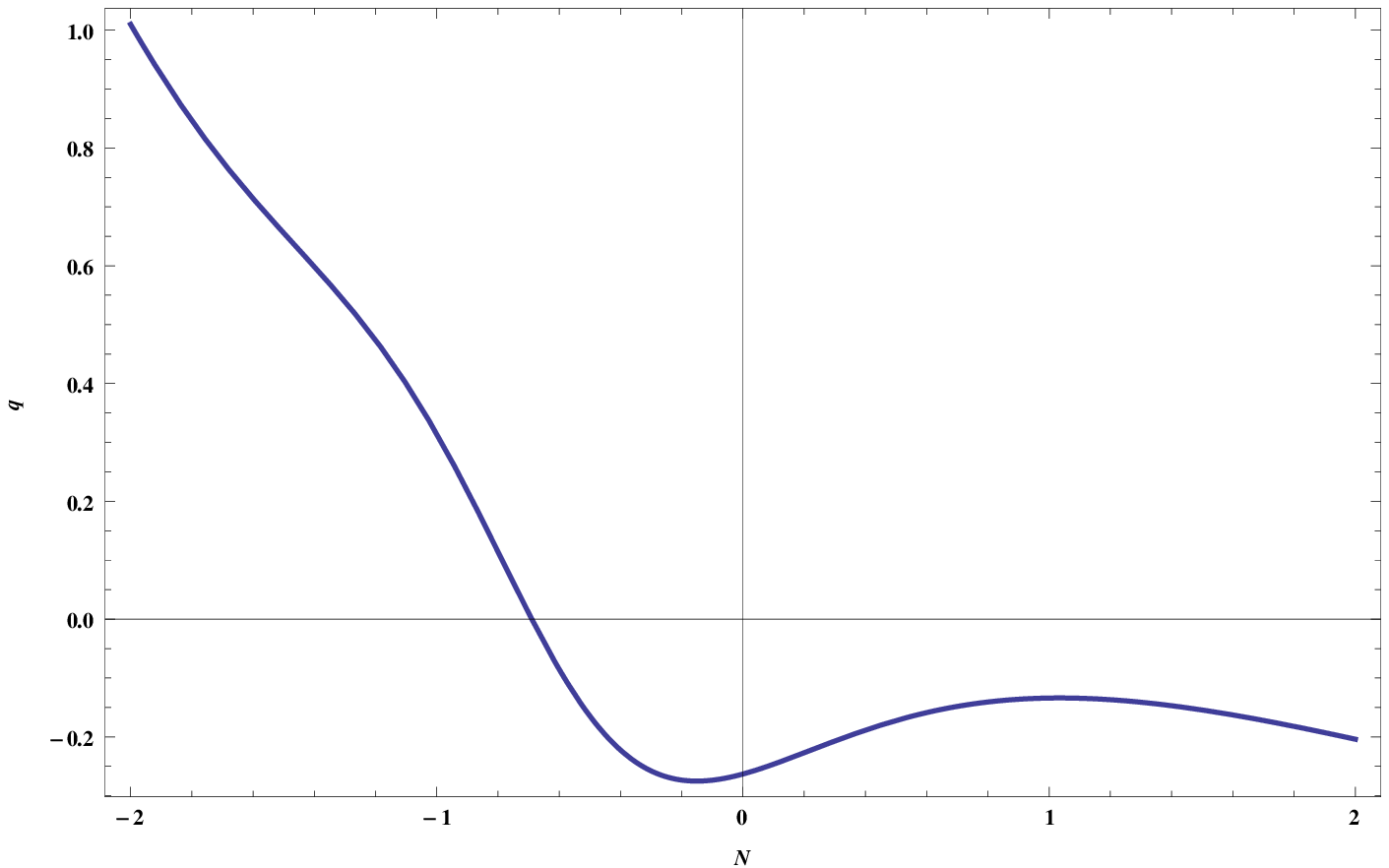}
 % sol1.eps: 0x0 pixel, 300dpi, 0.00x0.00 cm, bb=0 0 407 263
 \caption{Plot of deceleration parameter ``q'' against $N$ for $\lambda_0 = -1.5$ and $\delta_0 = 0.1$.}
\end{center}
 \end{figure}

\section{Discussion}

The stability of the various chameleon scalar field models are investigated in the context of the present accelerated expansion of the universe in the present work. The aim is not to suggest any new model of dark energy, but rather to look at the various possibilities where a chameleon field can indeed serve as driver of the acceleration, starting from a decelerated situation. An unstable fixed point might describe the initial stage of the universe from where a small perturbation could trigger the start of the evolution whereas a stable fixed point is apt to describe the final stage of the universe. Unlike many of the dark energy models, a chameleon field has the distinct possibilities of being detected and hence that of being nullified as well. Thus it warrants attention regarding the choice of the favoured combination of the dark energy potential and the coupling with matter. \\

For the sake of convenience, the functions $V$ and $f$ are classified as either exponential or not. So in all there are four such combinations, which are quite extensively studied in the present work. In fact any old (or new) chameleon model with given $V=V(\phi)$ and $f=f(\phi)$, the stability criteria need not be checked afresh. The present investigation provides a complete set of choices for $V(\phi)$ and $f(\phi)$ and can serve as the diagnostics. \\

 It is found that if both $V$ and $f$ are exponential functions,  $\delta$ and $\lambda$ are not dynamical variables but are rather some parameters. The stability of the fixed points depends on the values of these parameters and it opens up a possibility for a transient acceleration for the universe around the present epoch. The other observation is that when $V$ is exponential and $f$ is not, $V$ effectively resembles a cosmological constant and when $f$ is exponential but $V$ is not, the matter-chameleon coupling is actually broken. \\

It is interesting to note that the fixed point $n_3$ with $\lambda = 0$ and $p_1$, $q_1$ and $m_1$ are in fact the same, where there is no fluid and the scalar field sits at a non-zero minimum. So all the different combinations of $V$ and $f$ coincide at this fixed point which formally resembles a de Sitter model. \\

We can also find at least one example where the cosmological observations of the transition from the decelerated phase to the accelerated one can happen without any contradiction with the allowed band between the mass of the scalar field and its coupling with matter. \\

As already mentioned, the basic aim was not to propose a new chameleon model but rather to provide an exhaustive study of the stability characteristics of all possible combination of $V$ and $f$. This purpose is achieved quite comprehensively and as a bonus some interesting physical features are also noted. For example, it is noted that a chameleon field can give rise to a transient acceleration for the universe and certian combination of the potential $V$ and the coupling $f$ leads to a breakdown of the chameleon mechanism itself. \\

{\bf Acknowledgements:} One of the authors (N.R.) wishes to thank the CSIR (India) for financial support. The authors would like to thank Jayanta Bhattacharya for a really fruitful discussion. \\

 \end{document}